\documentclass[twocolumn, amssymb, amsmath,nobibnotes,nofootinbib, aps, pre,showpacs]{revtex4-2}
\usepackage{graphicx}
\usepackage{bm}
\usepackage{float}
\usepackage{dcolumn}% Align table columns on decimal point
\usepackage[hidelinks]{hyperref}
\usepackage[dvipsnames]{xcolor}%%%%%%%%%%%%%%%%%

\begin{document}
\title{Self organized criticality of magnetic avalanches in disordered ferrimagnetic material}
\author{Suman Mondal$^1$}
\author{Mintu Karmakar$^2$}
\author{Prabir Dutta$^3$}
\author{Saurav Giri$^1$}
\author{Subham Majumdar$^1$}
\email{sspsm2@iacs.res.in}
\author{Raja Paul$^2$}
\email{ssprp@iacs.res.in}
\affiliation{$^1$School of Physical Science, Indian Association for the Cultivation of Science, 2A \& B Raja S. C. Mullick Road, Jadavpur, Kolkata 700 032, INDIA}
\affiliation{$^2$School of Mathematical \& Computational Sciences, Indian Association for the Cultivation of Science, 2A \& B Raja S. C. Mullick Road, Jadavpur, Kolkata 700 032, INDIA}
\address{$^3$Jawaharlal Nehru Centre for Advanced Scientific Research,Jakkur, Bangalore, 560064, India}

\begin{abstract}
We observe multiple step-like jumps in a Dy-Fe-Ga based ferrimagnetic alloy in its magnetic hysteresis curve at 2 K. The observed jumps are found to have a stochastic character with respect to their magnitude and the critical field of occurrence, and the jumps do not show any temporal effect. The distribution of jump size follows a power law variation indicating the scale invariance nature of the jumps. We have invoked a simple two-dimensional random bond Ising type spin system to model the dynamics. Our computational work can qualitatively reproduce the jumps and their scale invariant character. It also elucidates that the flipping of antiferromagnetically coupled Dy and Fe clusters is responsible for the observed discrete avalanche-like features in the hysteresis loop. These characteristics indicate that the present phenomenon can be well described within the realm of self-organized criticality. 

\end{abstract}

\maketitle

%%%%%%%%%%%%%%%%%%%%%%%%%%%%%%%%%%%%%%%%%%%%%%%%%%%%%%%%%%%%%%%%%%%%%%%%%%%%%%%%%%%
\section{Introduction}

In physics and materials science, many systems, when driven by a slowly varying external parameter, can show avalanches in their physical properties. They include flux penetration in type-II superconductor~\cite{sc}, Barkhausen noise due to the domain distribution in ferromagnet~\cite{barkhausen0}, earthquake~\cite{earth_quake}, sand-piles~\cite{sandpiles}, forest fire~\cite{forrest_fire} and so on. One of the important characteristics of these systems is the scale invariance of the avalanche height~\cite{power_law}. The distribution of the jumps shows a power law behavior, $D(s) \sim s^{-\alpha}$, where $D(s)$ is the probability of an avalanche of height $s$, and $\alpha$ is an exponent mostly lying between 1 and 2. 

\par
The scale invariant avalanche dynamics of such systems are often described by the phenomenon of self-organized criticality (SOC)~\cite{SOC,BTW,Jensen,PBak}. The physical systems showing SOC are generally dissipative and locally interacting. The SOC organizes itself into self-organized metastable states, which transform from one to another via avalanches. Despite its complexity, the SOC has basic statistical features that are regulated by power laws\cite{PBak}. In the case of ferromagnetic systems, the magnetization shows a series of small jumps when slowly driven by a magnetic field and it is interpreted on the basis of SOC. This is called the Barkhausen effect, and it is related to the sudden reversal of the ferromagnetic domains~\cite{barkhausen}.

\par
Recently, few magnetically phase separated materials are reported to show multiple metamagnetic jumps under a varying magnetic field. The systems showing such jumps include various Mn-site doped manganites ~\cite{Mahendiran,Rana,Nair,Rana2,Liao,Ouyang,Tang}, Fe-site doped CeFe$_2$~\cite{SBR, AH}, Gd$_5$Ge$_4$~\cite{EM,JL} and its alloys. These jumps are characteristically different from the Barkhausen noise with a relatively larger variation of $M$. It is generally believed that these metamagnetic jumps occur due to the field-induced transition of an antiferromagnetic (AFM) cluster to a ferromagnetic (FM) one in those AFM/FM phase-separated systems. The jumps are often found to vary systematically with the sweep rate of $H$~\cite{Hardy,Suresh}. With an increasing sweep rate, the jump shifts to lower fields. This indicates {\it non-stationary} nature of the jumps. Even jumps are found to occur spontaneously, if one waits at a fixed temperature and magnetic field for a sufficient amount of time~\cite{hardy3,Suresh3}.

\par
The classical theory of SOC demands that a slowly driven system should have a stationary critical state devoid of any external fine tuning~\cite{barkhausen}. Therefore, the metamagnetic jumps in the phase-separated system cannot be labeled strictly as a classical SOC phenomenon. All the above systems are characterized by strong magneto-elastic coupling and the jumps are associated with the structural transition. Therefore, the internal strain at the interface of clusters plays an important role in the observed jumps.

\par 
The question remains, can there be a system showing large metamagnetic jumps obeying the SOC scenario? In the present work, we chose a relatively simpler system DyFe$_3$, which does not show any structural instability down to 4 K~\cite{neutron}. Here Dy and Fe sublattices are aligned antiparallel giving rise to a ferrimagnetic state. The Zeeman energy is supposed to be large due to the large moments at the Dy and Fe sites, which can facilitate spin flip under an applied magnetic field. We doped nonmagnetic Ga at the Fe site to introduce disorder in it. The magnetization curve of DyFe$_3$ is event-less showing small coercivity and saturation of moment above 15 kOe. On the other hand, the Ga-doped samples show clear ultrasharp metamagnetic jumps. This provides with us an opportunity to study the avalanches, where structural instability is unlikely to play a major role. 

\par 
In the present work, we have mostly focused on the sample, where 1/6 Fe is replaced by Ga (nominal composition DyFe$_{2.5}$Ga$_{0.5}$). The experimental data are supported by our classical Monte Carlo based simulation of the magnetic hysteresis loop. 

\section{Experimental details}
Polycrystalline samples of DyFe$_3$ and DyFe$_{2.5}$Ga$_{0.5}$ were prepared by standard argon arc melting technique and subsequent annealing. Structural characterization of the samples was performed by room temperature powder x-ray diffraction experiment using Cu-K$_{\alpha}$ radiation. We found that the parent DyFe$_3$ compound crystallizes in a rhombohedral structure, while Ga doped DyFe$_{2.5}$Ga$_{0.5}$ has a hexagonal structure (see fig.~\ref{xrd} of appendix~\ref{appendixA}). The dc magnetization ($M$) of the samples was measured using a Quantum Design SQUID magnetometer (MPMS3) as well as using the vibrating sample magnetometer of Quantum Design Physical Properties Measurement System (PPMS). The resistivity ($\rho$) and the magnetoresistance were measured by four-point technique using the same PPMS. 
%%%%%%%%%%%%%%%%%%%%%%%%%%%%%%%%%%%%%%%%%%%%%%%%%%%%%%%%%
\section{Experimental Results}

\begin{figure}[t]
\centering
\includegraphics[width = 8 cm]{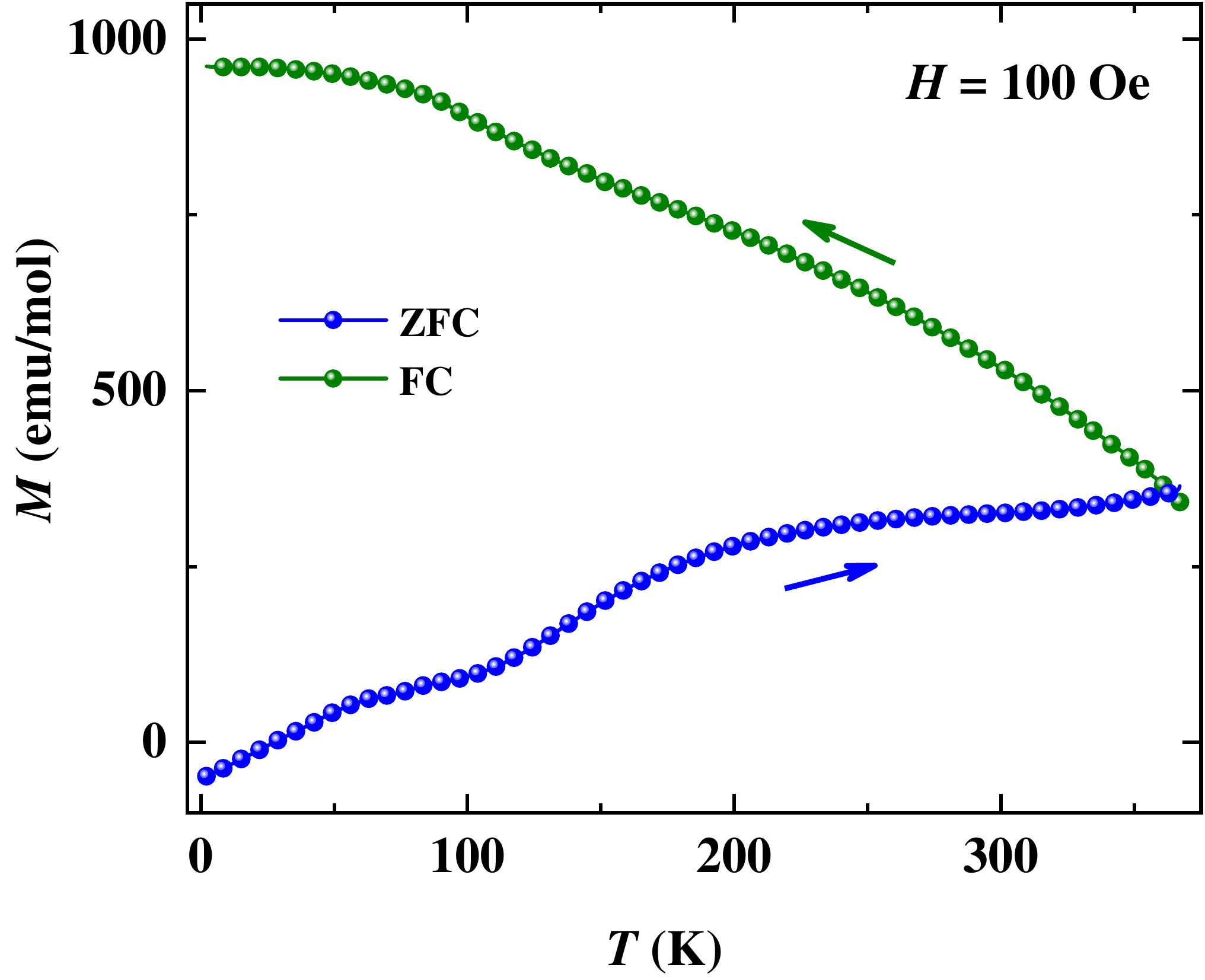}
\caption{Field-cooled (FC) and zero-field-cooled (ZFC) magnetization ($M$) data as a function of temperature ($T$) measured under 100 Oe of magnetic field for DyFe$_{2.5}$Ga$_{0.5}$.}
\label{mag}
\end{figure}

\subsection{Magnetization}

Fig.~\ref{mag} depicts the temperature ($T$) variation of $M$ of DyFe$_{2.5}$Ga$_{0.5}$ recorded in zero field cooled (ZFC) and field cooled (FC) protocols under 100 Oe of magnetic field. The ferrimagnetic nature of the undoped DyFe$_3$ is reported previously by Plusa {\it et al.} with ordering temperature and compensation point of 615 K and 525 K respectively~\cite{Plusa}. Ga doping in the Fe site is expected to reduce the magnetic ordering temperature. For DyFe$_{2.5}$Ga$_{0.5}$, we see large irreversibility between FC and ZFC curves, which extends up to the maximum temperature of measurement (370 K). This signifies that the magnetic ordering temperature is at least above 370 K. The FC-ZFC irreversibility indicates the presence of disorder in the system. 
\par
\begin{figure}[t]
\centering
\includegraphics[width = 8 cm]{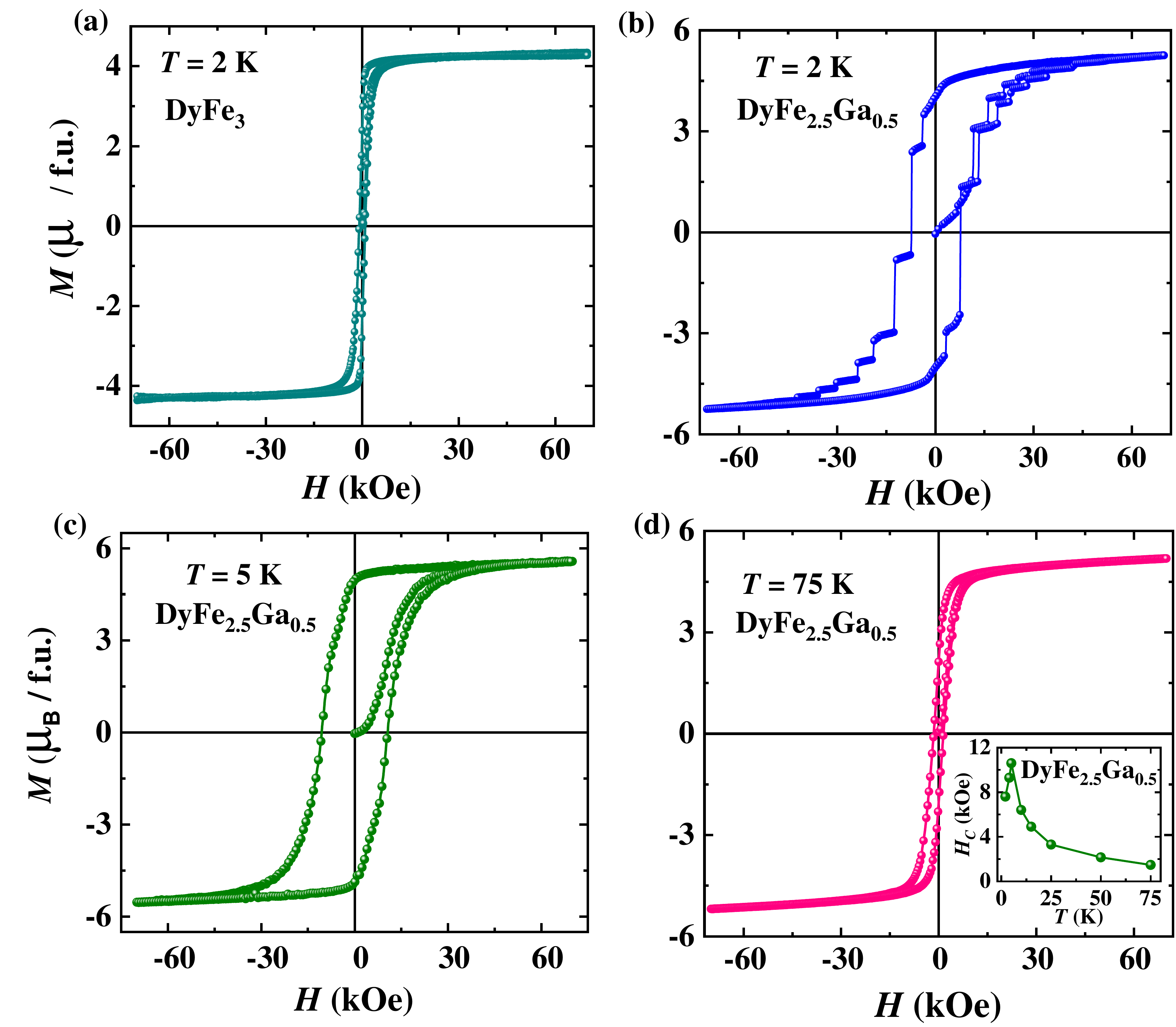}
\caption{(a) to (d) show full magnetization loop recorded at different temperatures for DyFe$_{2.5}$Ga$_{0.5}$. The inset depicts the temperature variation of the coercive field.}
\label{mh}
\end{figure}

Fig.~\ref{mh} (a) and (b) show the isothermal variation of $M$ with magnetic field ($H$) of DyFe$_3$ and DyFe$_{2.5}$Ga$_{0.5}$ samples at 2 K respectively. $M$ versus $H$ curve for the undoped DyFe$_3$ at 2K [fig.~\ref{mh} (a)] rises sharply at low fields and saturates beyond 10 kOe with a coercive field of 920 Oe. The saturation moment is found to be 4.3 $\mu_B$. This moment arises from the antiparallel arrangements of Dy and Fe spins~\cite{neutron,RFe3}. 
\par
The most striking observation of the present work is found in the 2 K isotherm of DyFe$_{2.5}$Ga$_{0.5}$ [fig.~\ref{mh} (b)]. We find successive jumps in $M$ as the field is swept between $\pm$ 70 kOe. The jumps are sharp and they are present in all five legs (including the virgin leg). $M$ eventually saturates above 60 kOe of field with saturation moment $m_{sat}$ = 5.3 $\mu_B$. Here $m_{sat}$ is higher than the undoped sample because a part of Fe is replaced by nonmagnetic Ga. Since the total moment is $m_{sat} = m_{Dy} - m_{Fe}$ (Fe sublattice is antiparallel to Dy in the ferrimagnetic state), the reduction in Fe content will enhance $m_{sat}$. The coercive field of the Ga doped sample is much higher ($\sim$ 7.5 kOe) than the parent one. Notably, the occurrence of multiple jumps vanishes as the temperature is slightly increased. The isotherm at 5 K [fig.~\ref{mh} (c)] shows a smooth variation of $M$ with $H$. A similar smooth isotherm is also observed at 75 K [fig.~\ref{mh} (d)], albeit with the much lower value of the coercive field. The inset of fig.~\ref{mh} (d) shows the temperature variation of the coercive field ($H_{coer}$) recorded between 2 and 75 K. $H_{coer}$ shows a non-monotonous variation with $T$ with a peak around 7 K. In disordered granular systems, such kind of variation of $H_{coer}$ is common~\cite{Fiorani}. 

\par
\begin{figure}[t]
\centering
\includegraphics[width = 9 cm]{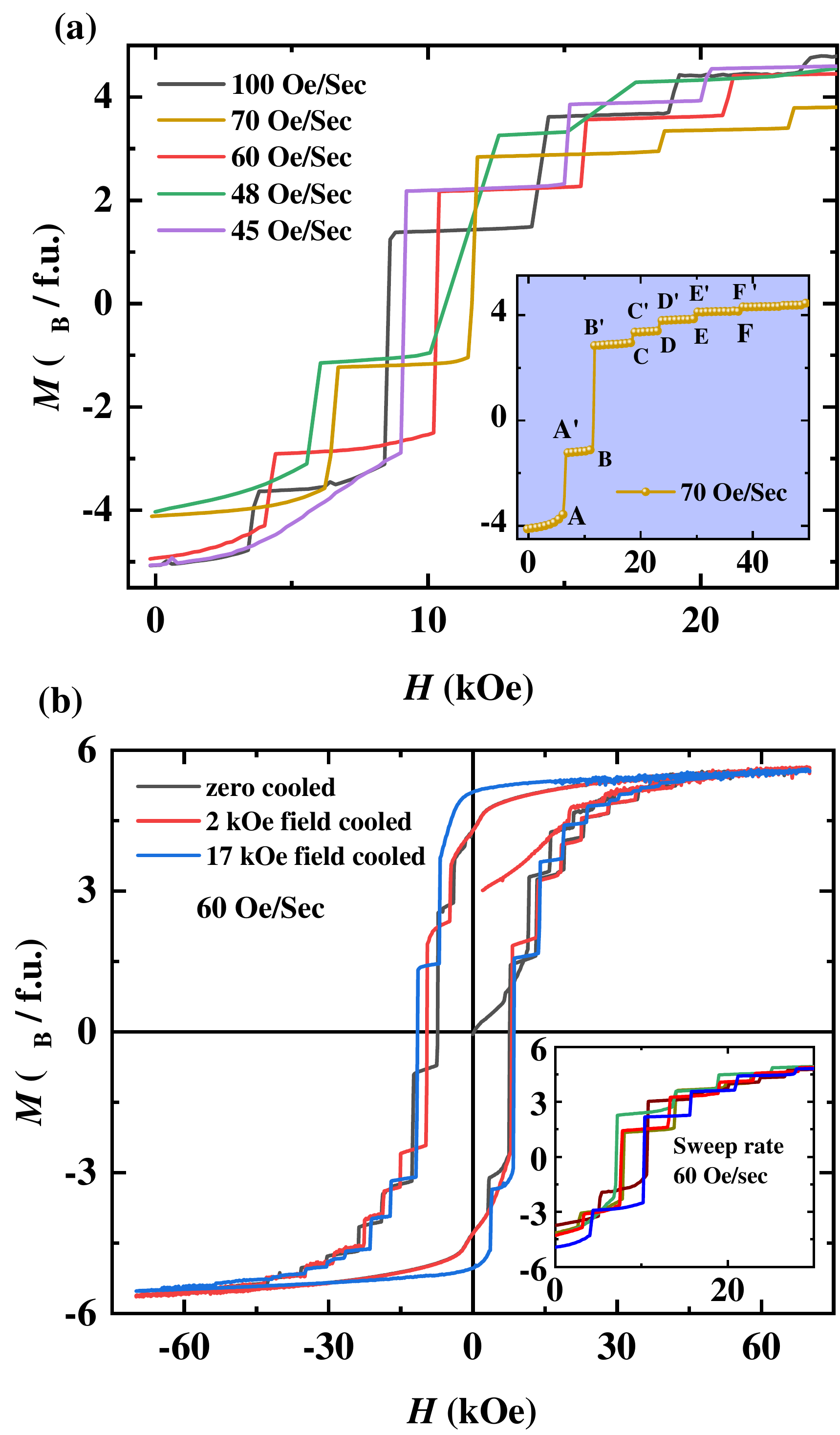}
\caption{(a) shows the fifth leg (0 to 70 kOe in the first quadrant, after returning from $-$70 kOe) of the magnetization loop recorded at 2 K for different sweep rates of the magnetic field ($H$). (b) represents the full loop at 2 K after the sample being cooled in different magnetic fields. Inset of (b) depicts the fifth leg of the hysteresis loop recorded at 2 K for different cycles.}
\label{jumps}
\end{figure}

We also recorded $M-H$ curves for DyFe$_{2.5}$Ga$_{0.5}$ at different sweep rates ($\dot{H}$ = 100, 70, 60, 48, and 45 Oes$^{-1}$) in the first quadrant, after the sample being returned from $-$70 kOe of field [fig.~\ref{jumps}(a)]. It is interesting to note that the position of the jumps does not vary systematically with the sweep rate. Even for several consecutive runs at $\dot{H}$ = 60 Oes$^{-1}$, the jump fields are markedly different (see the inset of fig.~\ref{jumps} (b)). Nevertheless, the observed jumps follow a certain pattern for all values of $\dot{H}$, and this is illustrated by the run at $\dot{H}$ = 70 Oes$^{-1}$ (see the inset of fig.~\ref{jumps}(a)). There is a common pattern in the jumps denoted by $AA^{\prime}$, $BB^{\prime}$, $CC^{\prime}$, $DD^{\prime}$..., with $AA^{\prime}< BB^{\prime}$, $BB^{\prime}>CC^{\prime}$, $CC^{\prime} \sim DD^{\prime}$. Although the positions, $A$, $B$, $C$, $D$.. are random on the $H$-axis, they occur in the same order and arrangement. Unlike previously reported multiple metamagnetic jumps~\cite{Hardy,Suresh,hardy3,Suresh3}, DyFe$_{2.5}$Ga$_{0.5}$ do not show a systematic sweep rate dependence. 

\par
We also studied the effect of the cooling field ($H_{cool}$) on the hysteresis loop. For that purpose, the $M-H$ isotherm is recorded at 2 K, after the sample being cooled from 300 K under a certain $H_{cool}$. The shift in the hysteresis loop on field cooling is called the exchange bias effect and it is expressed in terms of $H_{eb} =(H_+ + H_-)/2$ H$_+$ and H$_-$ stand for the positive and negative intercepts of the magnetization curve with the field axis respectively). For DyFe$_{2.5}$Ga$_{0.5}$, we see an increase of the magnitude of $H_-$, keeping $H_+$ (loop spreads in the negative side of $H$ asymmetrically). It is difficult to say whether such a shift arises from exchange bias or the shift of a jump. The jumps maintain the same pattern, but their height and magnitude vary randomly with $H_{cool}$.

%%%%%%%%%%%%%%%%%%%%%%%%%%%%%%%%%%%%%%%%%%%%%%%%%%%%%%%%%%
\par
%%%%%%%%%%%%%%%%%%%%%%%%%%%%%%%%%%%%%%%%%%%%%%%%%%%%%%%

%%%%%%%%%%%%%%%%%%%%%%%%%%%%%%%%%%%%%%%%%%%%%%%%%%%%%%%

%%%%%%%%%%%%%%%%%%%%%%%%%%%%%%%%%%%%%%%%%%%%%%%%%%%%%%%%%%
\par
\begin{figure}[t]
\centering
\includegraphics[width = 8.5 cm]{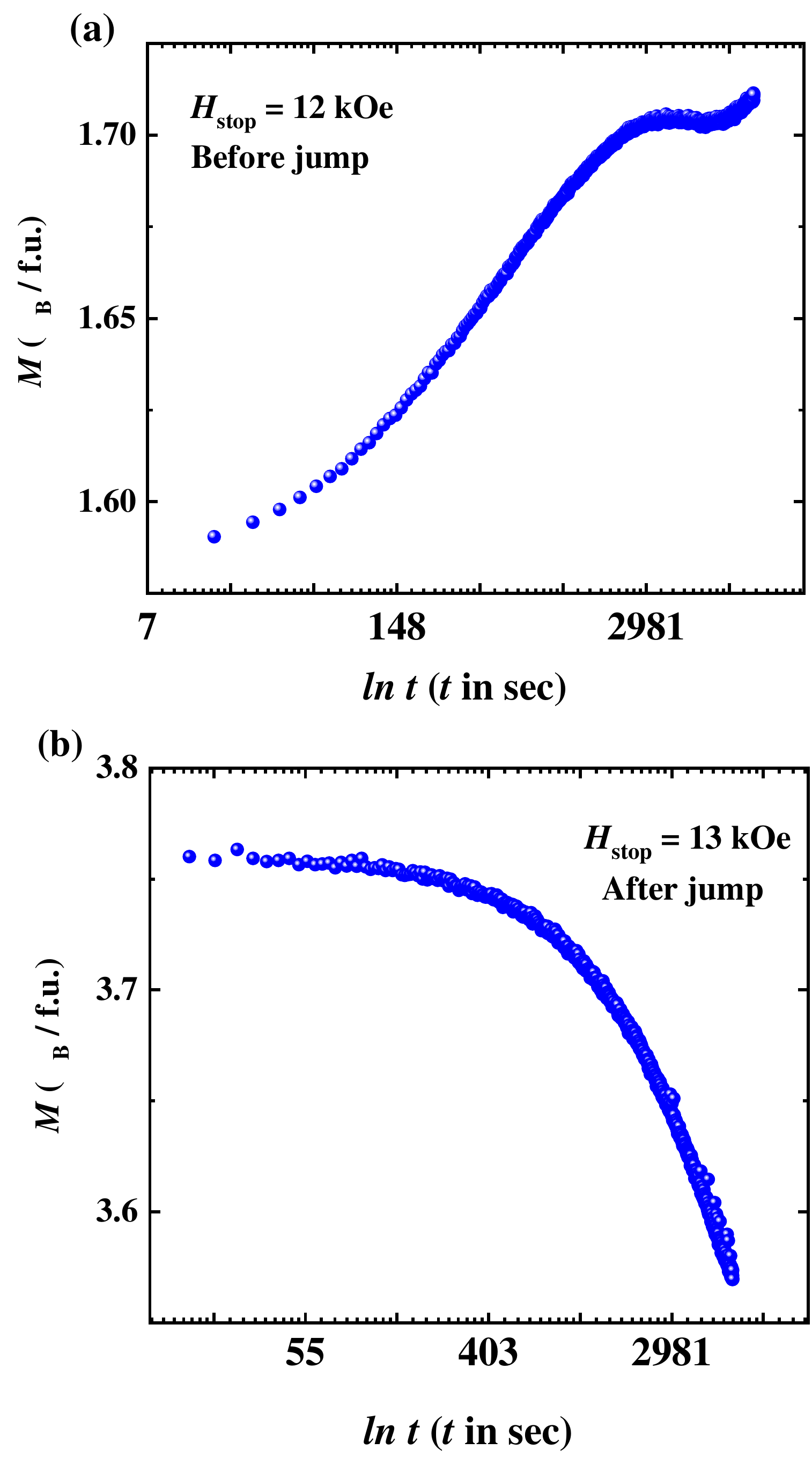}
\caption{Magnetization as a function of time for different protocols measured at 2 K after the application of a magnetic field of (a) 12 kOe; (b) 13 kOe of field. Before the measurement, the sample was first cooled down to 2 K, and it was kept at this temperature for the rest of the measurements.}
\label{relax}
\end{figure}

Temporal effects are often associated with magnetization jumps. It is found that the jump can spontaneously occur if one waits for sufficient time at a point on the $M-H$ loop at a field slightly lower than the critical field for a particular jump~\cite{hardy3,Suresh3}. We studied this phenomena by noting the variation of $M$ with time ($t$) in DyFe$_{2.5}$Ga$_{0.5}$. In fig.~\ref{relax} (a), we have shown the relaxation ($M$ vs $t$ data) after applying $H_s$ = 12 kOe of field ($H_s$ slightly lower than the field required for BB$^{\prime}$ jump). However, no spontaneous jump is seen even after waiting for 180 minutes, though $M$ changes by 8\% in 40 minutes. In fig.~\ref{relax} (b), measurement was performed at $H_s$ = 13 kOe, where the BB$^{\prime}$ jump already took place. Interestingly, $M$ drops with $t$, and a 5\% relaxation is seen. 
\par
Multiple metamagnetic jumps are common among glassy magnetic systems~\cite{Suresh,JK,MA}. To confirm the glassy nature in DyFe$_{2.5}$Ga$_{0.5}$, we measured the filled-cooled-field-stop memory effect (not shown here). The sample was cooled from 300 K under $H$ = 100 Oe with intermediate zero-field stops at $T_i$ (= 200, 150, 100 and 50 K) for 60 minutes. On reaching 2 K, the sample was heated back to 300 K in the presence of $H$ = 100 Oe. However, no feature is observed at the stopping temperatures $T_i$ during heating. This rules out the possibility of a glassy magnetic state in the system~\cite{mem1,mem2}.

\begin{figure}[t]
\centering
\includegraphics[width = 8 cm]{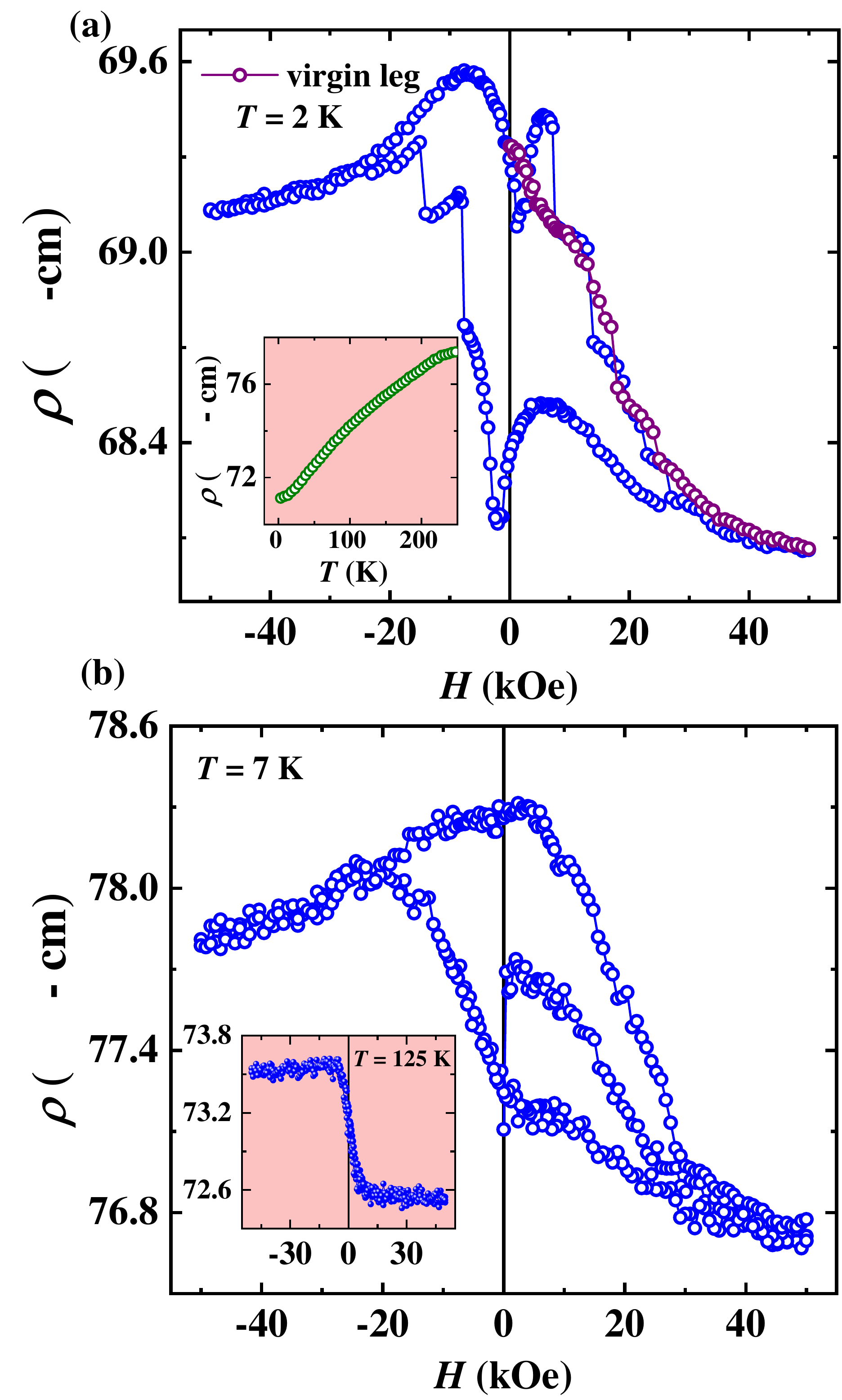}
\caption{(a) and (b) respectively show the variation of electrical resistivity ($\rho$) as a function of the field at 2 and 7 K. The inset of (a) represents the variation of electrical resistivity as a function of temperature. The inset of (b) shows the variation of $\rho$ as a function of $H$ at 125 K}
\label{mr}
\end{figure}
\par
%%%%%%%%%%%%%%%%%%%%%%%%%%%%%%%%%%%%%%%%%%%%%%%%%%%%%%%%%%

\subsection{Resistivity}
DyFe$_{2.5}$Ga$_{0.5}$ shows metallic behavior in the $T$ dependent plot of resistivity ($\rho$) (see inset of fig.~\ref{mr}(a)). In figs.~\ref{mr} (a) and (b), we have plotted the $H$ variation of $\rho$ measured at 2 and 7 K respectively. In the 2 K data $\rho(H)$ data, the ultra-sharp jumps are also present, and they approximately correspond to the similar critical fields for the jump as observed in the magnetization data. Similar to the $M(H)$ data, the jumps are absent at a higher temperature of 7 K. In both 2 and 7 K data, the full five quadrants $\rho(H)$ curves (between $\pm$ 50 kOe) form a hysteresis loop. Such observation indicates that the electronic property of DyFe$_{2.5}$Ga$_{0.5}$ is intimately correlated with the magnetic state of the system. At 125 K (inset of fig.~\ref{mr} (b)), there is no loop and $\rho(H)$ saturates above 10 kOe showing small magnetoresistance.

\begin{figure*}[t]
\centering
\includegraphics[width = 14 cm]{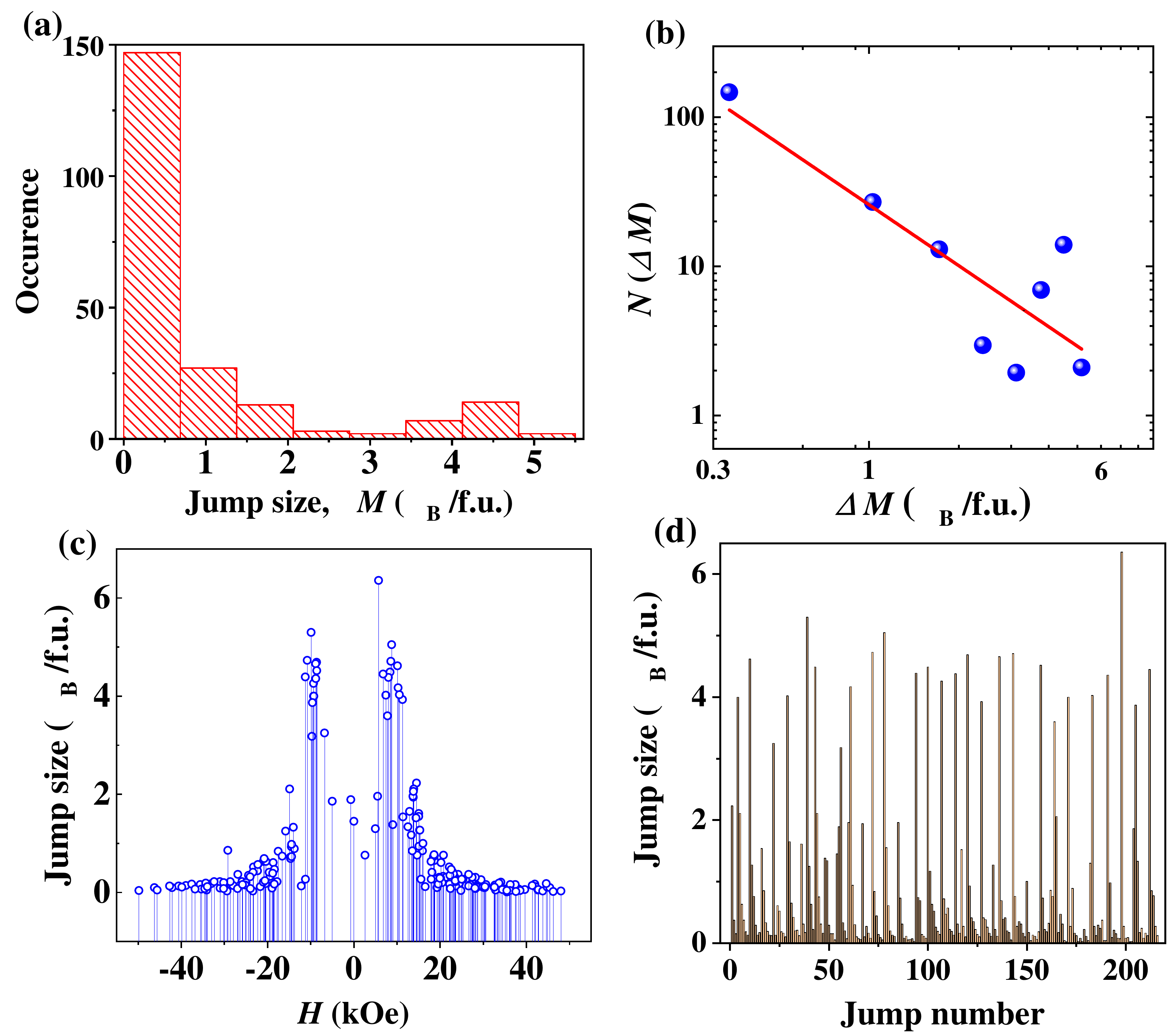}
\caption{(a) Histogram of the observed jumps in the experimental magnetization versus field data. (b) Number of occurrences as a function of jump size in $\log-\log$ scale. (c) Jump size as a function of the applied field. (d) Individual jumps are plotted as a function of the jump number.}
\label{jump_stat}
\end{figure*}

\subsection{Distribution of the jumps observed in experiment}
In fig.~\ref{jump_stat} (a), we have plotted the histogram of the jumps using our magnetization data depicted in fig.~\ref{jumps} (a). Since we do not find any correlation between the sweep rate and the jump size, we used data of all sweep rates to construct the histogram. We have used Sturge's rule ($\kappa = 1+3.22\log\nu$, $\kappa$, $\nu$ is the total number of data points) to calculate the number of bins, $\kappa$~\cite{HA}. Clearly, the smaller jumps ($\Delta M$ is low) are larger in number. In fig.~\ref{jumps} (b), we have plotted the number of occurrences $N(\Delta M)$ as a function of jump size. This provided a power law distribution, $N \sim \Delta M^{-\alpha}$ with $\alpha =$ 1.4.

\par 
In fig.~\ref{jump_stat} (c), we have studied the variation of jump size with the $H$. It is seen that the bigger jumps are found around 10 kOe and the jump size is smaller at higher fields [~\ref{jump_stat} (c)]. This is understandable from the isotherms recorded at 2 K. In ~\ref{jump_stat} (d), we have plotted the jump number as a function of jump size. We took twelve five-leg $M-H$ loops measured at 2 K, and serially counted the jumps for all the loops. A jump number is assigned to each jump. We find that the jump size span across multiple ranges, e.g., big jumps (4-5 $\mu_B$/f.u.), medium jumps (1-2 $\mu_B$/f.u.), and small jumps (below 0.25 $\mu_B$/f.u.) and the data correlates with the characteristics of fig.~\ref{jump_stat} (b). 

\section{Theoretical Model}
A vast majority of real-world materials contain impurities that introduce disorder in the system~\cite{Chaikin,Sethna}. For such systems, the central challenge is to predict large-scale behavior from local dynamics. Developing a corresponding model of interacting spins originates from the inspection of the experimental data.
%~\cite{BinderK}. Even the simplest models have so far eluded analytic explanation that they can only be examined statistically~\cite{Binder}. 
The experimental observation of magnetization jumps in the Dy-Fe-Ga alloys are found to be scale invariant and stationary~\cite{barkhausen,stationary1,stationary2}. This prompted us to model the system for quantitative analysis of the hysteresis loops. We simulated the jumps with changing magnetic fields and quantify their distribution. 

\begin{figure*}[hbt!]
\includegraphics[width=\columnwidth]{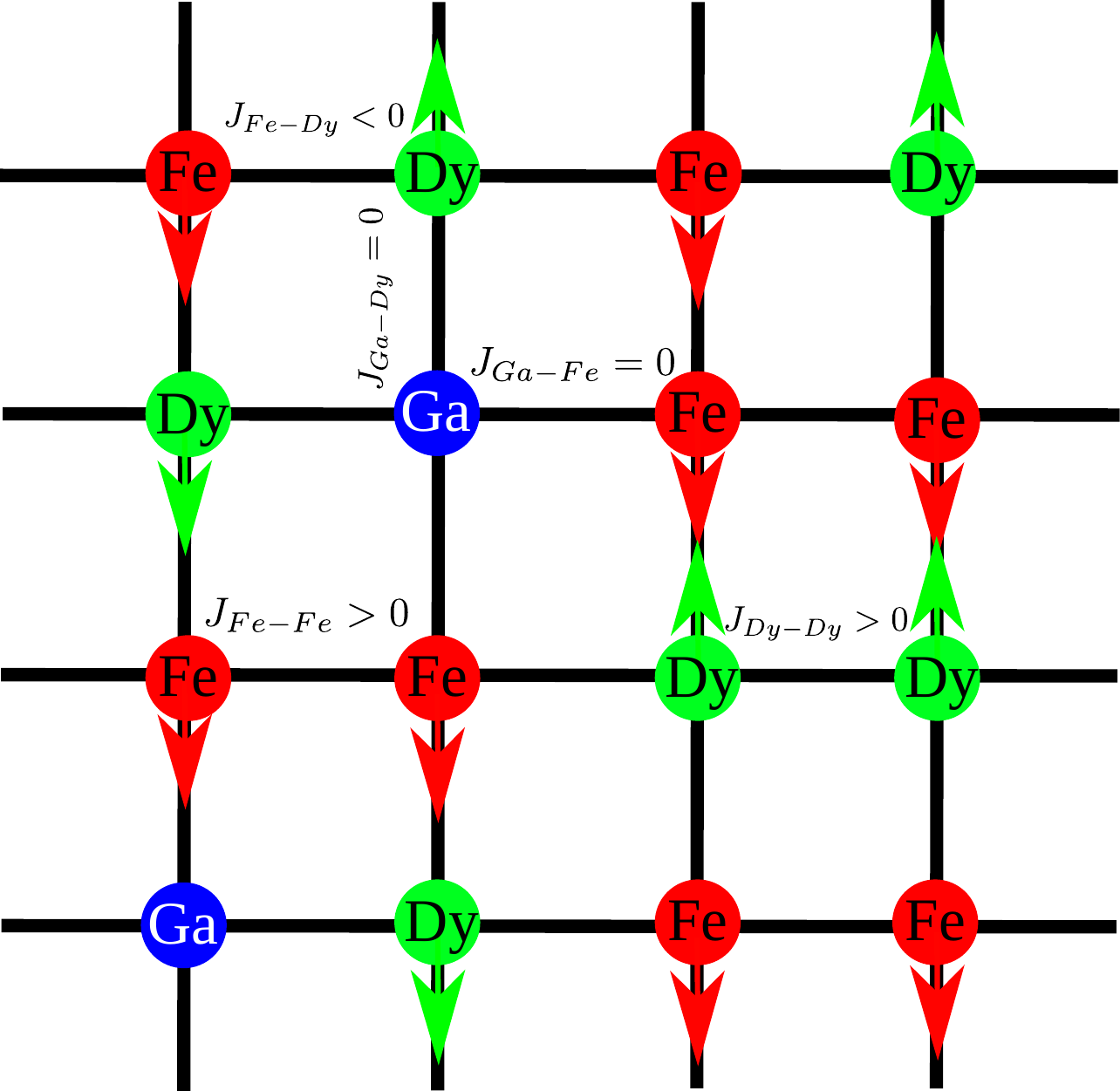}% Here is how to import EPS art
\caption{\label{fig:Model_fig} Model Schematic shows a lattice structure in two-dimensions with Fe and Dy randomly distributed in a $3:1$ ratio. Identical and non-identical atoms are linked via Ferromagnetic and anti-ferromagnetic interactions, respectively. Disordered is introduced by nonmagnetic Ga replacing Fe according to fraction $f_{Ga}$. %Field ($h$) is applied along the down spin direction.
}
\end{figure*}

We choose the generalized random-bond Ising model. The Hamiltonian is defined as;

\begin{equation}
\label{freeenergy}
\mathcal{H}(h)=-\sum_{i,j}\mathcal{J}_{ij}\sigma_i \sigma_j-h\sum_i \sigma_i-h\sum_{i\in Dy} L_i^{Dy}
\end{equation}
where $\sigma_i=\pm$ 1 are the local spin moment, $\mathcal{J}_{ij}$ are the nearest-neighbor interactions, and $h$ is the external field. Two different types of spins corresponding to Fe and Dy are assumed on a square lattice in $3:1$ ratio, respectively. The spin value $\sigma_i$ is taken unity for both Dy and Fe because their spin moments are almost equal~\cite{Jensen2}. The direction of the external magnetic field is fixed along one of the spin variables (e.g. $-$ve or $+$ve). $L_i^{Dy}=1$ is the orbital moment for Dy alone, directed along Dy's spin moment, and contributes to the Hamiltonian at specific lattice sites occupied by Dy atoms. The nearest-neighbor interactions in the first term of eqn.~\ref{freeenergy} can be negative or positive ({\it i.e.,} AFM or FM respectively) depending upon the type of interacting particles. Ferromagnetic interaction between two Fe is given by FM coupling $\mathcal{J}_{ij}$ = 1, while between two Dy it is $\mathcal{J}_{ij}$ = 0.05 (see fig.~\ref{fig:Model_fig}). Interaction between Fe and Dy is AFM and coupling constant $\mathcal{J}_{ij} =-$2. The $\mathcal{J}_{ij}$ values are obtained from the relative strength of the magnetic interaction available in the literature~\cite{duc1,duc2}. $f_{Ga}$ defines the fraction of the site disorder due to the doping of non-magnetic Ga at the expense of Fe. For the experimentally studied samples DyFe$_3$ and DyFe$_{2.5}$Ga$_{0.5}$, $f_{Ga}$ is 0 and 0.2 respectively.

%The free energy Eq. (\ref{freeenergy}) corresponds to the simple ferromagnetic ($J_{ij}>0$) and antiferromagnetic ($J_{ij}<0$) Ising model. For $h=0$ Eq. (\ref{freeenergy}) corresponds to spin-glass model. The dynamics of the system described by Eq. (\ref{freeenergy}) has been extensively studied as a function of dimensionality and disorder type\cite{Sethna,Binder}. 
\section{Simulation Details:} 
An ensemble of Ising spins is considered on a two-dimensional ($2D$) square lattice. Initially, the system is prepared with Fe and Dy randomly distributed at $3:1$ ratio on a $64\times 64$ lattice represented by the model described in fig.~\ref{fig:Model_fig}. The disorder is introduced by replacing randomly chosen Fe sites by Ga with fraction $f_{Ga}$. The system is then equilibrated using Glauber dynamics, which is a computer simulation of the Ising model (a magnetism model). Using Monte Carlo simulation without the external field, at $T = 0$, only the sign of the energy differences is required for the Glauber dynamics. We have simulated this system with periodic boundary condition. The initial configuration is random and single spin flip Glauber dynamics has been used for subsequent updating, {\it i.e.}, a spin is picked up at random and flipped if the resulting configuration has lower energy, never flipped if the energy is raised and flipped with probability $1/2$ if there is no change in energy on flipping. Next, a small external magnetic field ($h$) is applied and the system is re-equilibrated. Note that the notations used to represent the external magnetic field and average magnetization in the simulation are to mark a quantitative difference with the experimental values. The external field is raised slowly and the system is equilibrated for every value of $h$. Average magnetization ($M$) is computed as a function of $h$. The external field increases until average magnetization saturates. The process is repeated by decreasing $h$. The simulation yields a single hysteresis loop as a function of model parameters, $(f_{Ga},h,\mathcal{J}_{ij})$.

\section{Numerical Results}
\begin{figure*}[hbt!]
\includegraphics[width = 15 cm]{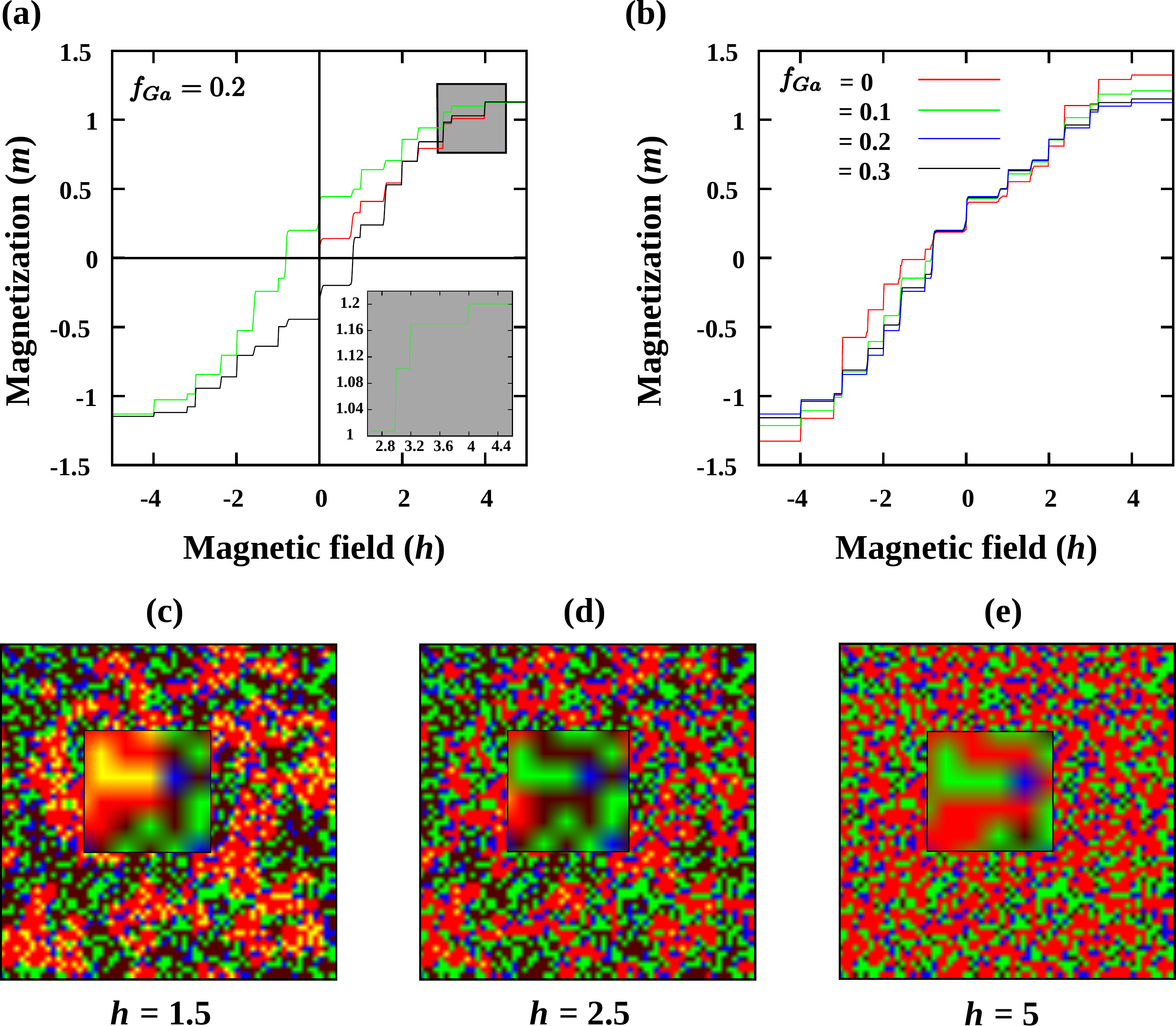}% Here is how to import EPS art
\caption{\label{fig:drawing} (a) Hysteresis loop of random-bond Ising model for disorder fraction $f_{Ga} =$ 0.2. Inset: Many small jumps occur at a large field in the hysteresis loops. (b) Magnetization profiles for four distinct disorder fractions $f_{Ga}=$0, 0.1, 0.2, and 0.3 are illustrated when the external field ($h$) is reduced. (c), (d), and (e) are snapshots of spin configurations before and after large jumps in the magnetization at field strengths of $h =$ 1.5, 2.5, and $h =$ 5 (as in figure (a)), respectively. Red, brown, green, yellow, and blue represent Fe($+$), Fe($-$), Dy($+$), Dy($-$), and non-magnetic Ga, respectively. The magnetic field ($h$) is applied in the $+ve$ direction. The snapshots are magnified in the middle to provide a detailed insight into cluster flipping.}
\end{figure*}

%Now, ferromagnetic interaction is always helpful in restoring order to the Ising system in response to an external field. The number of antiferromagnetic interactions between Fe and Dy in the system is considerable in the absence of disorder through the spin-glass state. As a result, the sharp change in magnetization with respect to the external field is gradual.
The initial system is prepared with zero external magnetic field, and then it is subjected to gradually increasing field $h$ in step of 0.02. For each field value, the system is equilibrated up to 10$^5$ Monte Carlo steps. Fig.~\ref{fig:drawing} (a) shows the simulated hysteresis loop for $f_{Ga} =$ 0.2, which clearly demonstrates multiple steps in magnetization. The data show intermittent jumps ($\Delta m>$ 0) and stationary phases ($\Delta m$ = 0) while the external field is increased. A similar characteristic is observed when the external magnetic field is decreased. Fig.~\ref{fig:drawing}(b) compares magnetization profile for three different disorder fractions $f_{Ga} =$ 0, $f_{Ga} =$ 0.1, $f_{Ga} =$ 0.2 and $f_{Ga} =$ 0.3. Jumps observed in the magnetization are a signature of frustration in spin-spin interaction, usually characteristics of a glassy magnetic system~\cite{Fischer,Rieger, Van_Hemmen}. It is to be noted that jumps are present even for $f_{Ga} =$ 0, where there is no doping at the Fe sites. This is because the present model is inherently disordered due to the random occupancy of Fe and Ga in 3:1 ratio. The nature of the jumps remains nearly unchanged with $f_{Ga}$. 

\par
Snapshots at different field strengths are shown in figs. ~\ref{fig:drawing} (c) , (d) and (e) and figs.~\ref{fig:snap} (a), (b), (c), and (d) of appendix~\ref{appendixA} for the $f_{Ga} =$ 0.2 lattice. At $h$ = 0, we find the finite clusters of Dy and Fe with up($+$) and down($-$) spin configurations. With increasing $h$ ({\i.e.} in the $+$ direction), the clusters flip in the direction of the applied field, which produces jumps in the hysteresis curve.

\begin{figure*}[hbt!]
\includegraphics[width = 14 cm]{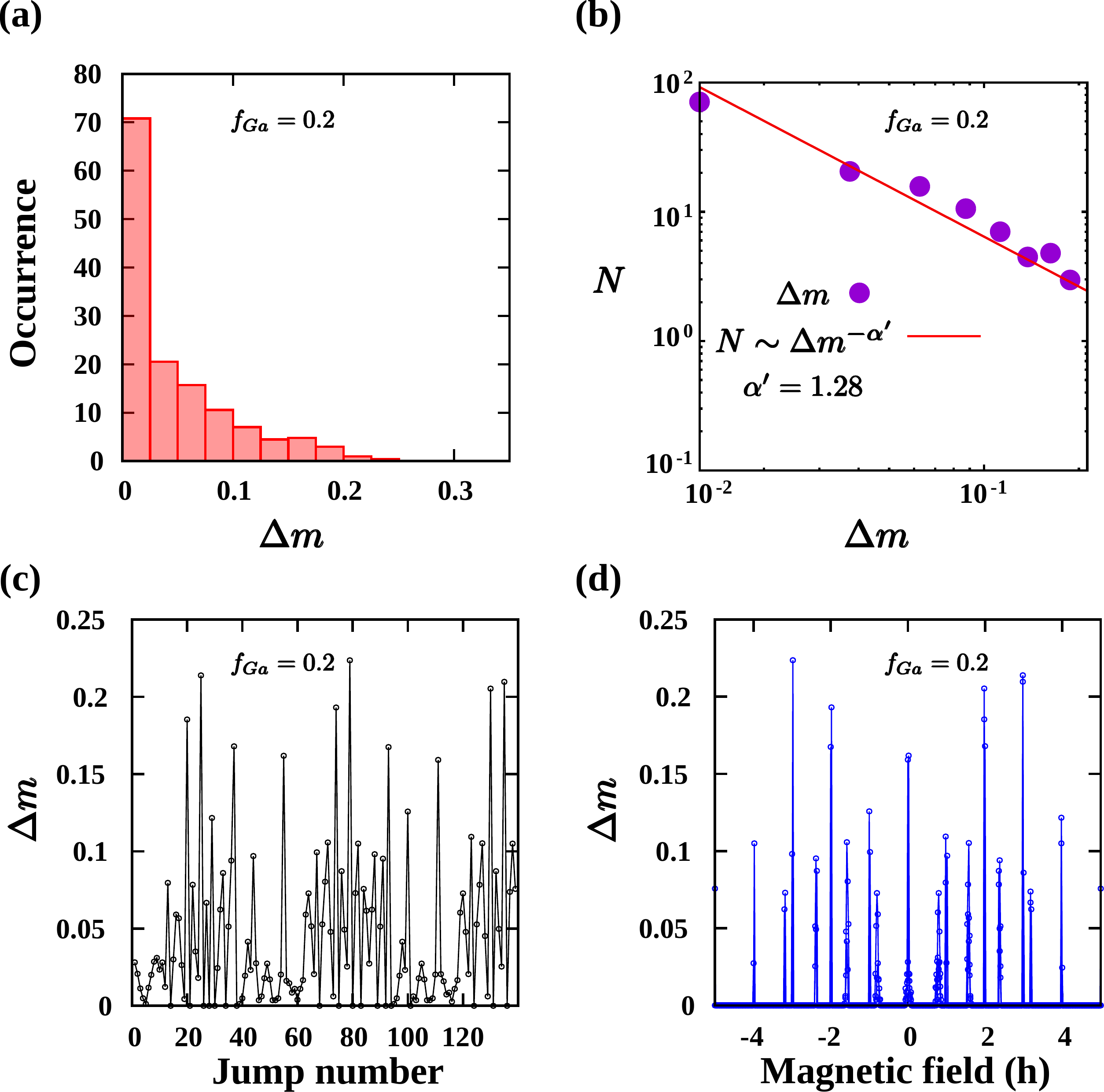}% Here is how to import EPS art
\caption{\label{fig:fitting} (a) Distribution of magnetization jumps showing occurrence ($N$) as a function of $\Delta m$ for disorder fraction $f_{Ga}=0.2$. The data points correspond to an average over $100$ ensembles. (b) The power-law distribution of $N$ vs. $\Delta m$ falls with exponent $\alpha^\prime=1.28$. (c) Magnetization jump sizes $(\Delta m)$ of a single hysteresis loop demonstrates jump sizes ranging from small to large. (d) Jump size is plotted as a function of the external field $h$.}
\end{figure*}

A statistical analysis of the jump size $\Delta m$ projects the probability distributions of avalanche sizes $N(\Delta m)$ for $f_{Ga} = $0.2 in fig.~\ref{fig:fitting}(a). The data points represent an average of more than 100 ensembles. A comparison of this distribution function with respect to the zero disorder scenario ($f_{Ga} = $0) shows that for both cases, the occurrence $\Delta m$ of for very small $\Delta m$ appears to be significantly higher than large $\Delta m$ (see figs.~\ref{fig:hist} (a) and (b) in appendix~\ref{appendixA}). Moreover, the histogram stretches to larger $\Delta m$ for the zero disorder. 

\par
The log-log plot of the distribution of magnetization for disorder fraction $f_{Ga}=$0.2 (as in fig.~\ref{fig:fitting}(a)) is shown in fig.~\ref{fig:fitting}(b). The graph indicates a power law distribution of $N \sim \Delta m^{-\alpha^\prime}$. Fitting the data points with the given expression yields the exponent $\alpha^\prime=$1.28. The value of $\alpha^\prime$ is found to be close to $\alpha$ obtained from the power law fitting of the experimental data (see fig.~\ref{jump_stat} (b)). The avalanche sizes $(\Delta m)$ of a single hysteresis loop are depicted in fig.~\ref{fig:fitting}(c). The jump sizes obtained from the hysteresis loop are plotted sequentially showing multiple regimes corresponding to small, intermediate, and large jumps of magnetization. The data signifies that the size of domain flips (avalanche) due to a marginal and systematic change in the external magnetic field is likely uncorrelated as obvious from the power law distribution in fig.~\ref{jump_stat} (b). We have further shown $\Delta m$ as a function of $h$. It is clear that for low fields, the magnetization jump is substantial. Due to the AFM interaction between Fe and Dy, two adjacent domains remain frustrated and merge to form a larger domain, lowering the net surface energy as the external field is increased. In contrast, jumps are small in a large field. Because the majority of the adjacent domains have already flipped along the field direction. So, with a large field, only a few small clusters or individual spins remain to flip along the field direction until saturation magnetization is reached.

\section{Discussion}
Our combined experimental and theoretical simulation reveal that the magnetic avalanches found in the Dy-Fe-Ga alloy is a manifestation of self-organized criticality. Experimentally, the magnetization jumps in the studied alloy are stochastic, do not show any systematic change with the change in the rate of the external driving parameter (here magnetic field), and most importantly, they are scale-invariant following a power law distribution. The Monte-Carlo simulation of a model 2D analog of the real material also supports our experimental data, where the change in the magnetization appears to be discontinuous as a function of the external field and obeys a power-law distribution. 
\par
Numerous shreds of evidence suggest the existence of power law distribution in Nature. Remarkably, in some cases, the exponents of the distribution are the same for systems with very different microscopic descriptions~\cite{Zipf}. SOC is attributed to the evolution of a complex system towards criticality in presence of local interaction. For Dy-Fe-Ga alloy, the local interaction is the magnetic correlations between Dy and Fe atoms. 
\par
In DyFe$_3$, Dy and Fe spins show AFM correlation and they are aligned antiparallel resulting in ferrimagnetism. Doping by some amount of non-magnetic element Ga in Fe site introduces disorder in the Fe sublattice. The finite jumps in the magnetization data indicate the flipping of magnetic clusters than the individual spins. Because of the disorder, the system is characterized by the coexistence of spontaneously ordered but oppositely oriented neighboring magnetic domains of Dy and Fe respectively. Because of the AFM interaction between Fe and Dy, two such adjacent domains remain frustrated and cannot merge to become a larger domain lowering the net surface energy. To grow a larger domain, one of the oppositely oriented domains must be flipped entirely, which can be achieved by the external magnetic field. However, due to frustration in the local interaction, a sufficiently strong external magnetic field equivalent to the surface area of the domain must be provided. Therefore, a slight change in the external magnetic field often does not alter the overall magnetization of the system. As a result, we observe a stair-case-like feature in the magnetization isotherms both in the experiment (fig.~\ref{jumps}) and in simulation (fig.~\ref{fig:drawing}).

\par
In figs.~\ref{fig:drawing}(c) and (d), snapshots of the spin clusters are shown before and after the jumps respectively. %Red, brown, green, yellow and blue respectively represent the Fe($+$), Fe($-$), Dy($+$), Dy($-$) and Ga clusters ($+$ and $-$ are moment directions parallel and anti-parallel to the field respectively). 
At vanishing external fields, Dy and Fe clusters assume both $+$ve and $-$ve orientations and remain frustrated due to antiferromagnetic interaction (see fig.~\ref{fig:snap} (a) of appendix~\ref{appendixA}). In this scenario, one recognizes that Dy($+$) spins are surrounded by Fe($-$) and vice versa. At the other extreme, when the external field is increased substantially in the $+$-direction, several Fe($-$) clusters flip toward the direction of the applied field (Fe($-$) $\rightarrow$ Fe($+$)) producing a large jump in the magnetization. This is evident from the increase of red domains of Fe($+$) spins at $h$ = 5. In this case, the Zeeman energy overcomes the frustration due to AFM coupling present between Dy and Fe clusters. Interestingly, a contrasting spin arrangement occurs at an intermediate field. The enlarged section of the snapshots in figs.~\ref{fig:drawing}(c), and (d), indicate some Dy($-$) clusters flip to Dy($+$) as the $h$ is increased from 1.5 to 2.5. This is accompanied by the flipping of Fe($+$) to Fe($-$) adjacent to the Dy. Such an arrangement is energetically favorable due to the large moment of Dy compared to Fe. The outcome is manifested by smaller jumps in the magnetization. The flipping events connect nearby domains of like spins. A cascade of spin-flip and domain rearrangement leads to the formation of bigger clusters when the Zeeman energy overcomes the interfacial AFM interaction between Fe and Dy clusters. Flipping of larger domains then gives rise to larger jumps in the magnetization for a small increase in the external magnetic field. At a very large external field, the flipping events are dominated by the external field and the Fe($-$) spins flip back to Fe($+$) (enlarged part of fig. ~\ref{fig:drawing} (e)) leading to the saturation of magnetization. Note that, considering a purely anti-parallel arrangement of Dy and Fe moments, the expected moment should be 4.30 $\mu_B$/f.u. in DyFe$_{2.5}$Ga$_{0.5}$. However, our experimental value of the saturation moment is 5.25 $\mu_B$/f.u. Such discrepancy is likely due to the parallel arrangement of some Dy and Fe clusters, which takes place through discrete jumps. 

In conclusion, the discrete magnetization jumps in the Dy-Fe-Ga compound is found to be a spectacular manifestation of self-organized criticality. Our work indicates the flipping of the finite domains of Fe and Dy sublattices in the otherwise antiferromagnetically coupled spin system. The theoretical analysis can broadly reproduce the experimental results. It should be kept in mind that the present computational analysis is a simplified approach, and predictions are qualitative. We considered a two-dimensional system and random distribution of the spins. The experiment, however, is carried out on a three-dimensional design. Moreover, the arrangement of the atoms is not random in the actual system. Therefore, the computational model devoid of the crystal structure of DyFe$_3$ contains an intrinsic disorder, unlike the real systems. Computational work on a larger system with an actual three-dimensional crystal lattice can help us better understand the phenomenon.

\section{Acknowledgment}
MK is thankful to Council of Scientific and Industrial Research (CSIR), India for financial support through Grant No. 09/080(1106)/2019-EMR-I. RP, MK thanks Indian Association for the Cultivation of Science (IACS) for the computational facility. PD would like to acknowledge DST-SERB for his NPDF fellowship (PDF/2017/001061).
\appendix

\begin{figure*}
\centering
\includegraphics[width = 11 cm]{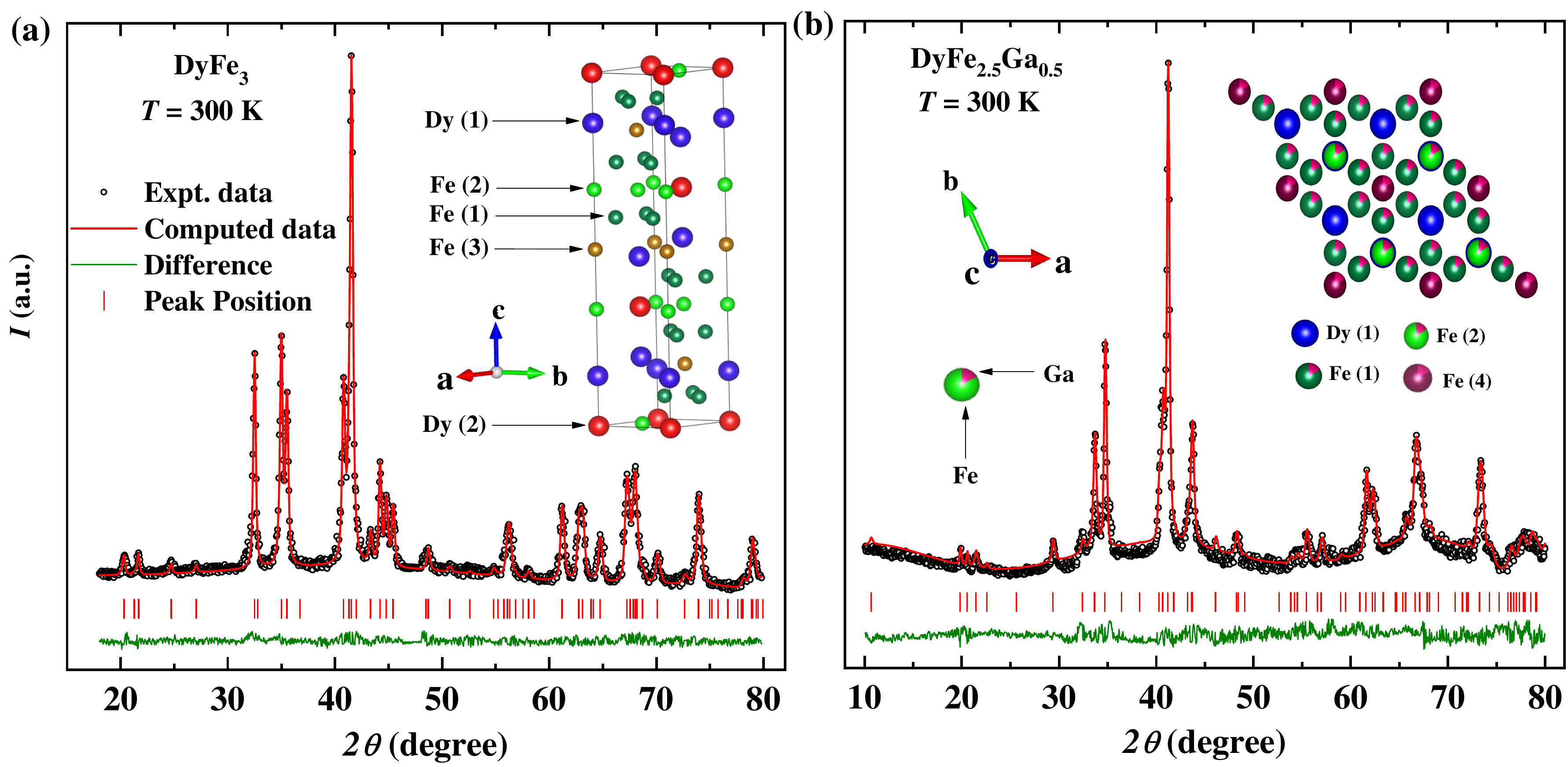}
\caption{(a) and (b) show powder X-ray diffraction (PXRD) patterns of DyFe$_3$ and DyFe$_{2.5}$Ga$_{0.5}$ sample respectively. The insets show the perspective view of the crystal structures.}
\label{xrd}
\end{figure*}

\section{CRYSTAL STRUCTURE}
\label{appendixA}
The X-ray diffraction patterns of the two samples are shown in fig.~\ref{xrd}. The undoped sample DyFe$_3$ crystallize in a PuNi$_3$-type rhombohedral structure (inset of fig.~\ref{xrd} (a)), with space group $R\bar{3}m$. In the literature, such structures are often represented as a hexagonal equivalent, i.e., $a = b \neq c$, $\alpha = \beta =$ 90$^{\circ}$, $\gamma =$ 120$^{\circ}$~\cite{buschow,chen,mk}. From our refinements, we get $a$ = 5.125~\AA and $c$ = 24.575~\AA. On the other hand, the Ga doped sample DyFe$_{2.5}$Ga$_{0.5}$ assumes a CeNi$_3$-type hexagonal structure with space group $P6_3/mmc$, where the refined lattice parameters are found to be $a$ = 5.165~\AA and $c$ = 16.560~\AA~\cite{RGa3}. The structure consists of hexagonal layers in the $a-b$ plane of the crystal as shown in the inset of fig.~\ref{xrd} (b). 

\begin{figure*}[hbt!]
\includegraphics[width= 16 cm]{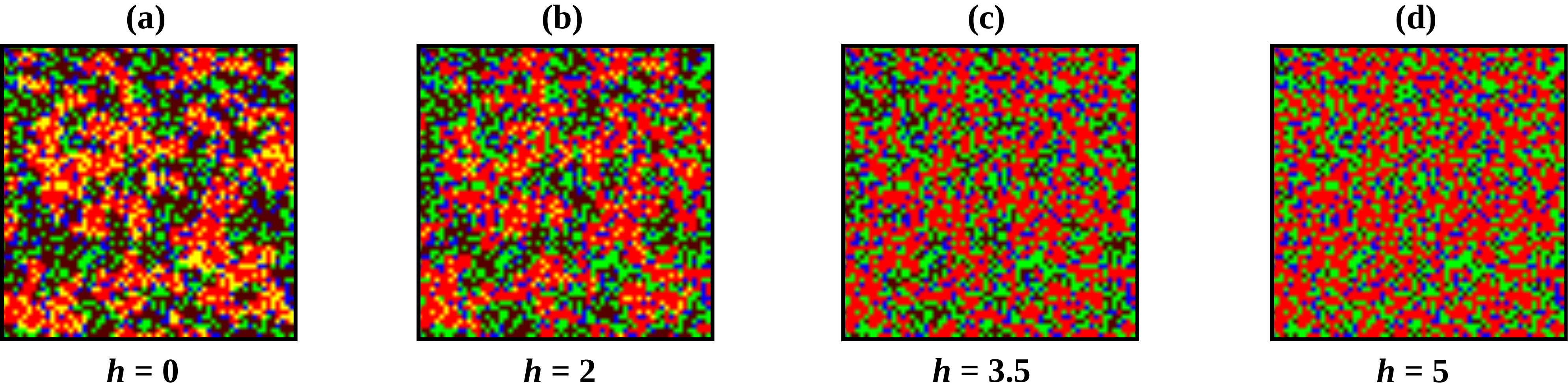}% Here is how to import EPS art
\caption{\label{fig:snap} Snapshots of the spin configurations Fe($+$), Fe($-$), Dy($+$), Dy($-$), and non-magnetic Ga is shown with increasing external field $h$ for disorder fraction $f_{Ga} = 0.2$.}
\end{figure*}

\begin{figure*}[hbt!]
\includegraphics[width= 12 cm]{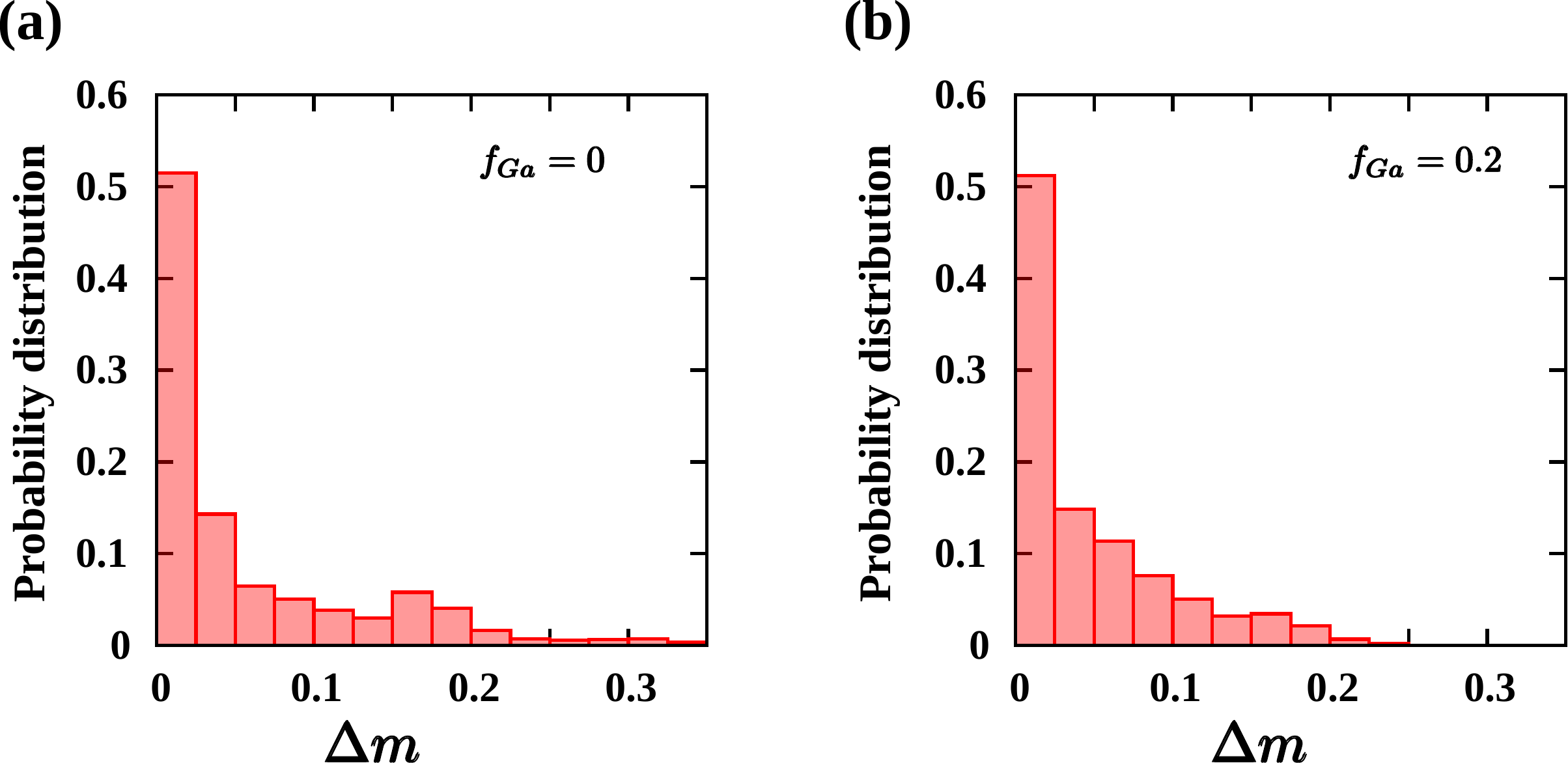}% Here is how to import EPS art
\caption{\label{fig:hist} Probability distributions of magnetization jump derived from the hysteresis loop with disorder fraction (a) $f_{Ga}=0,$ and (b) 0.2 respectively.}
\end{figure*}

\begin{figure*}[hbt!]
\includegraphics[width=12 cm]{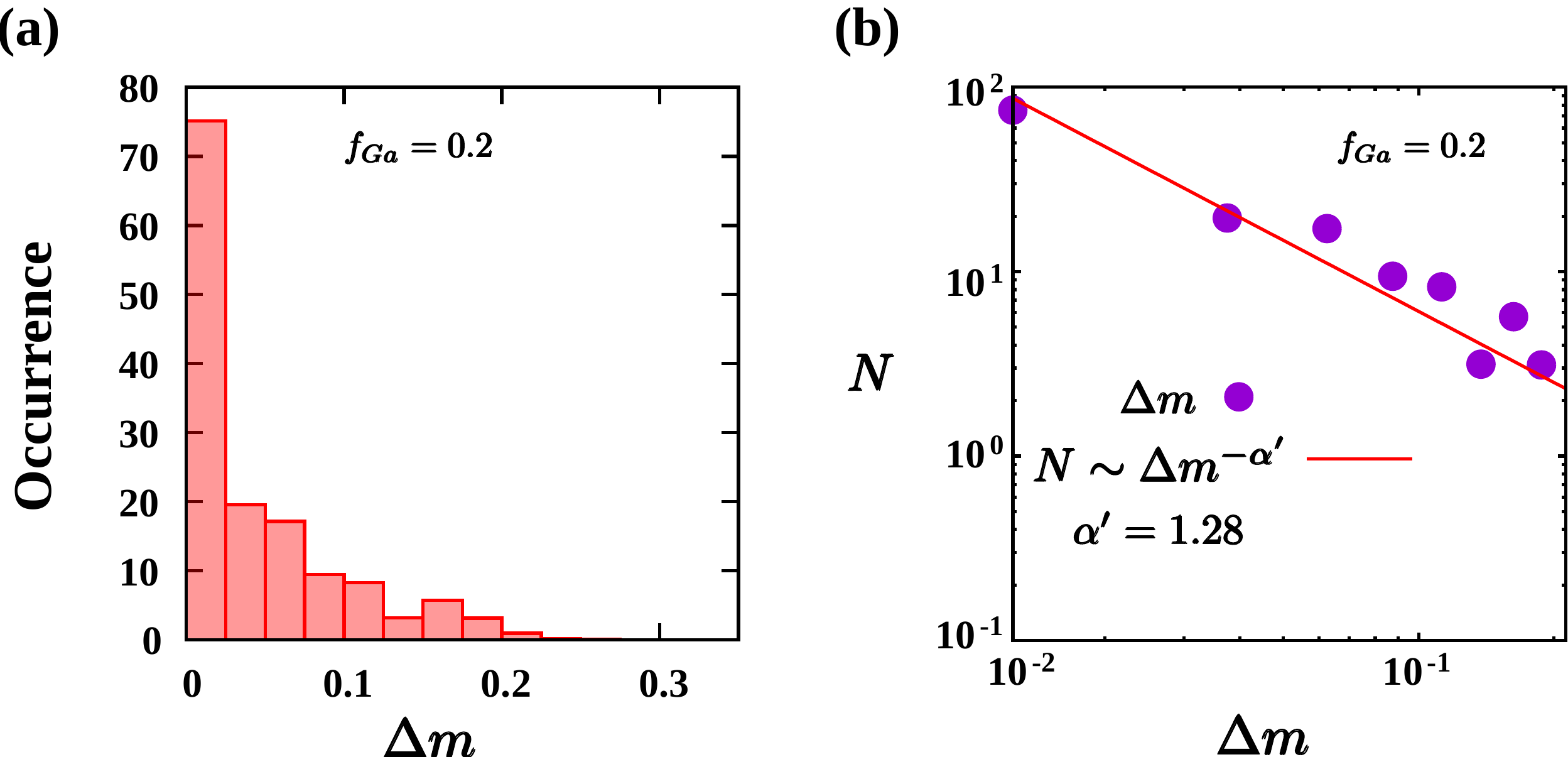}% Here is how to import EPS art
\caption{\label{fig:fitting_suppl} (a) Distribution of magnetization jumps exhibiting occurrence ($N$) as a function of $\Delta m$ for disorder fraction $f_{Ga}=0.2$ and system size $128$. The data points correspond to an average over $100$ ensembles. (b) The power-law distribution of $N$ vs. $\Delta m$ falls with exponent $\alpha^\prime=1.28$. }
\end{figure*}

\section{MAGNETIC MODELLING}
\label{appendixB}
In the case of Fe, there are 6 electrons in the 3$d$ orbital, so the total spin is $S_{Fe} =$ 2, while Dy has a total spin of $S_{Dy} =$ 5/2. Since the values are close, we considered the Fe and Dy spins are equal, and assigned the reduced spin value $\sigma_i = \pm$ 1 for both the elements. The Zeeman energy in the magnetic field $h$ for Dy is given by $\mathcal{H}_Z = -({\bf L} + g_0{\bf S})$~\cite{Jensen2}. Here ${\bf L}$ = Dy orbital moment, ${\bf S}$ = Dy spin moment and $g_0 =$ 2 is the Land\'e-$g$ factor for spin. We know for Dy, $L$ = 5 and $S$ = 5/2, and due to $g_0$, their contribution are the same in the Zeeman term. Therefore, in the Hamiltonian, we can take the reduced value of effective orbital contribution to be unity ($L_i^{Dy}=$1).

\end{document}